\documentclass[cameraready]{Interspeech}

\usepackage{graphicx}
\usepackage{subcaption}
\usepackage{multirow}
\usepackage[table]{xcolor} 
\usepackage{booktabs}      
\usepackage{threeparttable}

\usepackage{amsmath,amsfonts,bm}

\def\eqref#1{equation~\ref{#1}}

\def\1{\bm{1}}

\def\rmK{{\mathbf{K}}}

\def\rmQ{{\mathbf{Q}}}

\def\rmV{{\mathbf{V}}}

\DeclareMathAlphabet{\mathsfit}{\encodingdefault}{\sfdefault}{m}{sl}
\SetMathAlphabet{\mathsfit}{bold}{\encodingdefault}{\sfdefault}{bx}{n}

\def\gL{{\mathcal{L}}}

\usepackage{tikz}
\usepackage{array} 

\definecolor{colorT}{RGB}{0, 114, 178} 
\definecolor{colorM}{RGB}{213, 94, 0}   

\newcommand{\tikzbar}[1]{%
    \begin{tikzpicture}[baseline=0pt]
        \foreach \type [count=\i] in {#1} {
            \fill[color\type, draw=white, line width=0.5pt, rounded corners=0.5pt] (\i*0.8em, 0) rectangle (\i*0.8em + 0.8em, 0.8em);
        }
    \end{tikzpicture}%
}

\newcommand{\algo}{MamTra}

\title{\algo: A Hybrid Mamba-Transformer Backbone for Speech Synthesis}

\author[affiliation={1}]{Tan Dat}{Nguyen} 
\author[affiliation={1}]{Sangmin}{Bae} 
\author[affiliation={1}]{Joon Son}{Chung} 
\author[affiliation={2}, correspondingauthor]{Ji-Hoon}{Kim} 

\address{
    $^1$ Korea Advanced Institute of Science and Technology, South Korea \\
    $^2$ Chung-Ang University, South Korea
}

\email{tandat.kaist@kaist.ac.kr}


\usepackage{comment}

\begin{document}

\maketitle

\begin{abstract}
Despite the remarkable quality of LLM-based text-to-speech systems, their reliance on autoregressive Transformers leads to quadratic computational complexity, which severely limits practical applications.
Linear-time alternatives, notably Mamba, offer a potential remedy; however, they often sacrifice the global context essential for expressive synthesis.
In this paper, we propose MamTra, an interleaved Mamba-Transformer framework designed to leverage the advantages of Mamba’s efficiency and Transformers’ modeling capability.
We also introduce novel knowledge transfer strategies to distill insights from a pretrained Transformer into our hybrid architecture, thereby bypassing the prohibitive costs of training from scratch. 
Systematic experiments identify the optimal hybrid configuration, and demonstrate that MamTra reduces inference VRAM usage by up to 34\% without compromising speech fidelity--even trained on only 2\% of the original training dataset. Audio samples are available at\footnote{\href{https://mm.kaist.ac.kr/projects/mamtra/}{Project page: https://mm.kaist.ac.kr/projects/mamtra/}}.\looseness=-1

\end{abstract}

\section{Introduction}
The recent surge of Large Language Model (LLM)-based Text-to-Speech (TTS) systems has yielded near-human naturalness in expressive and multi-speaker speech generation~\cite{borsos2023audiolm, wang2023neural, du2024cosyvoice, chen2025f5, song2025distar, du2024cosyvoice2}. 
However, this success remains fundamentally dependent on autoregressive Transformer backbones. While their core self-attention mechanism effectively models global dependencies across the entire input sequences, it inherently incurs quadratic time and memory complexity with respect to sequence length~\cite{vaswani2017attention}.
In long-form synthesis scenarios, including podcasts, audiobooks, and streaming dialogue agents, this quadratic scaling leads to high latency and large key–value (KV) cache memory, making deployment impractical on edge devices~\cite{park2024long,bae2025hybrid}.\looseness=-1
\begin{figure}[t]
    \centering
    \includegraphics[width=0.94\linewidth]{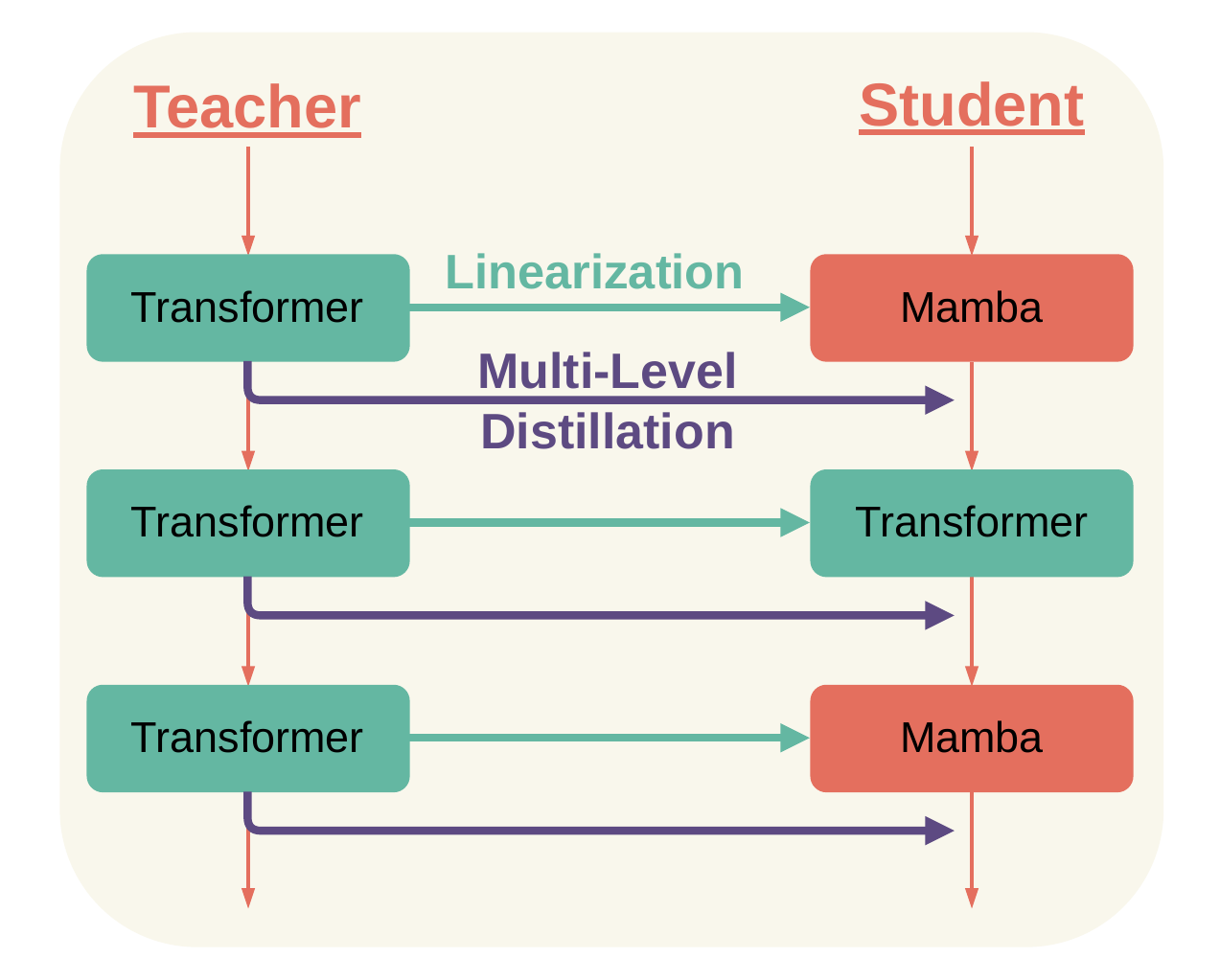}
    \caption{Overview of the hybrid Mamba-Transformer configurations for speech synthesis. Selective Transformer-to-Mamba layer transfer reduces training cost and accelerates convergence, while performance is recovered via knowledge distillation using less than 2\% of the teacher’s English training data.\looseness=-1}
    \label{fig:hybrid_model_distillation}
    \vspace{-3ex}
\end{figure}

Efforts to mitigate the computational bottlenecks of self-attention have spanned both architectural modifications—such as grouped-query attention (GQA)~\cite{ainslie2023gqa}, sparse attention~\cite{liu2025deepseek}, and pruning~\cite{ma2023llm, nguyen2025spade}—and inference optimizations like KV cache compression~\cite{ge2023model, zhang2023h2o} and speculative decoding~\cite{nguyen2025accelerating,li2025fast}. While effective, these methods fundamentally remain optimizations atop the quadratic attention mechanism rather than structural replacements. Consequently, \textit{linear} attention mechanisms have emerged as a scalable alternative computational primitive~\cite{sun2023retentive, gu2021efficiently, dao2024transformers, de2024griffin, yang2024gated,gu2024mamba}. In this context, State Space Models (SSMs)—most notably Mamba~\cite{dao2024transformers,gu2024mamba}—have gained particular traction for their ability to combine linear-time inference with strong local temporal modeling. This potential has driven recent adoption in a range of speech tasks, where Mamba-based architectures are demonstrating significant efficiency gains~\cite{Katharopoulos2020transformers,fu2023hungry,park2024long,yang2024gla,jiang2025speech,zhang2025mamba,gao2024speech,kwak2025ednet}.
However, this efficiency often comes at the cost of expressivity: pure SSMs still underperform Transformers at scale~\cite{jelassi2024repeat}, particularly in in-context learning~\cite{park2024can, grazzi2024mamba}, and long-context reasoning~\cite{ye2025longmamba}.\looseness=-1

To bridge this gap, \textit{hybrid} architectures have become a promising solution, strategically interleaving Mamba blocks with Transformer blocks. These designs combine the computational efficiency of Mamba for local acoustic modeling with the global semantic reasoning of Transformers~\cite{park2024can, lieber2024jamba, ren2024samba, dong2024hymba, bae2025hybrid}.
However, within the TTS domain, such hybrid designs remain largely unexplored. The \textit{only} existing attempt~\cite{zyphra2025zonos} relies on inefficient pretraining from scratch, and its technical implementation details have not been released.
To prevent the high cost of pretraining, we leverage insights from recent model conversion techniques~\cite{bick2024transformers,wang2024mambainllama, bick2025llamba, nguyen2025spade}. Specifically, we hypothesize that a pre-trained Transformer can be transformed into a hybrid TTS model by replacing a subset of its blocks with Mamba layers via attention-to-SSM parameter transfer.
This structured initialization ensures that the model retains teacher-level speech fidelity while achieving significantly faster inference.\looseness=-1

Our contributions are summarized as follows: \textbf{(1)} We introduce \textbf{\algo}, the 
first systematic study of hybrid Mamba-Transformer architecture for speech synthesis. It reuses pretrained Transformer weights to initialize both retained Transformer layers and newly introduced Mamba blocks, enabling selective replacement without pretraining from scratch (Figure~\ref{fig:hybrid_model_distillation}). 
\textbf{(2)} We conduct a speech-centric analysis of the hybrid design space, identifying where and how many to place Mamba layers and quantifying a clear efficiency–quality trade-off. Our results demonstrate a strong trade-off: using knowledge distillation on just 2\% of the English data compared to baseline yields a 34\% reduction in GPU memory with only a 0.25\% absolute increase in WER while maintaining the naturalness and expressiveness of the teacher.
\textbf{(3)} Finally, we validate MamTra's efficiency through both extensive experiments and theoretical analysis, confirming sub-quadratic complexity and sub-linear KV-cache growth. This corresponds to a reduction of up to 1.4$e^{11}$ FLOPs per token at a context length of 2,048.\looseness=-1

\begin{table}[t]
\centering
\caption{Architectural replacement strategies mapping Transformer (\textcolor{colorT}{blue blocks}) and Mamba (\textcolor{colorM}{orange blocks}) layers. {*}\,{BlockBeg} stands for {BlockBeginning}.}
\label{tab:strategies}
\begin{threeparttable} 
\begin{tabular}{llc}
\toprule
\textbf{Category} & \textbf{Strategy} & \textbf{Structure} \\ 
\midrule
\multirow{2}{*}{{Interleaved}} 
& BlockBeg{*}   & \tikzbar{T,M,M,T,M,M} \\ 
& BlockEnd   & \tikzbar{M,M,T,M,M,T} \\
\midrule
\multirow{4}{*}{{Contiguous}}  
& Front      & \tikzbar{M,M,M,T,T,T} \\
& Middle     & \tikzbar{T,T,M,M,T,T} \\
& Back       & \tikzbar{T,T,T,M,M,M} \\
& Sandwich   & \tikzbar{M,M,T,T,M,M} \\ 
\midrule
{Data-Driven}                  
& Importance & \tikzbar{T,M,T,T,M,T} \\ 
\bottomrule
\end{tabular}
\end{threeparttable}
\vspace{-3ex}
\end{table}

\section{MamTra Hybrid Architecture}

\subsection{Preliminaries}

\textbf{Transformer}~\cite{vaswani2017attention} has become the dominant backbone for sequence modeling due to its ability to capture long-range dependencies through self-attention mechanism. 
Each token attends to relevant context (e.g., preceding tokens in autoregressive models) by computing attention scores:\looseness=-1
\begin{equation}
\label{eq:attn}
\operatorname{Attention}(\rmQ,\rmK,\rmV) = \operatorname{softmax}\left( \frac {\rmQ\rmK^{\top}}{\sqrt{d_k}} \right)\rmV,
\end{equation}
where $\mathbf{Q}, \mathbf{K}, \mathbf{V} \in \mathbb{R}^{L \times d}$ are the query, key, and value matrices for sequence length $L$ and hidden dimension $d$. While caching the full KV history ensures precise global reasoning, its quadratic complexity $\mathcal{O}(L^2)$ severely bottlenecks long-form generation~\cite{vaswani2017attention}.\looseness=-1

In contrast, \textbf{Mamba}~\cite{ dao2024transformers,gu2024mamba} builds upon SSMs, which map a 1D input sequence $x(t)$ to an output $y(t)$ through a latent state $h(t)$. Unlike traditional time-invariant SSMs, Mamba introduces \textit{selectivity} by making the state parameters functions of the current input $\mathbf{x}_t$. The sequence is modeled through a compressed hidden state $\mathbf{h}_t \in \mathbb{R}^N$, updated \textit{recurrently}:\looseness=-1
\begin{equation}
\mathbf{h}_t = \mathbf{\overline{A}}_t \mathbf{h}_{t-1} + \mathbf{\overline{B}}_t \mathbf{x}_t, \quad \mathbf{y}_t = \mathbf{C}_t \mathbf{h}_t,
    \label{eq:ssm_discrete}
\end{equation}
where $\mathbf{\overline{A}}_t$ and $\mathbf{\overline{B}}_t$ are obtained by discretizing the continuous system matrices via a zero-order hold (ZOH)~\cite{gu2021efficiently} and the Euler method. By projecting the discretization parameters $\Delta_t$, $\mathbf{B}_t$, and $\mathbf{C}_t$ from the input $\mathbf{x}_t$, the model achieves a selective mechanism that dynamically propagates or forgets information based on the context~\cite{gu2024mamba}. Mamba leverages parallel scan algorithms~\cite{blelloch1990prefix, smith2023simplified} for highly efficient parallel training, and ensures linear $\mathcal{O}(L)$ decoding time with constant $\mathcal{O}(1)$ memory footprint during inference.\looseness=-1

\subsection{Hybridization Strategy}

We explore various Transformer-to-Mamba replacement strategies to balance global context modeling with local temporal efficiency. As summarized in Table~\ref{tab:strategies}, we categorize these into four design spaces: (1) \textbf{Interleaved} (BlockBeg\,/\,End), preserving periodic Transformer layers; (2) \textbf{Contiguous} (Front\,/\,Middle\,/\,Back\,/\,Sandwich), grouping layer types spatially into continuous blocks or at the model's boundaries; and (3) \textbf{Data-driven} (Importance)~\cite{wang2024mambainllama}, replacing less critical layers based on cosine similarity or WER criteria. Each strategy is evaluated across Transformer-Mamba ratios ranging from 1:1 to an aggressive 1:11 to determine the optimal functional regions for each primitive.\looseness=-1

\begin{figure}[t]
    \centering
    \includegraphics[width=\linewidth]{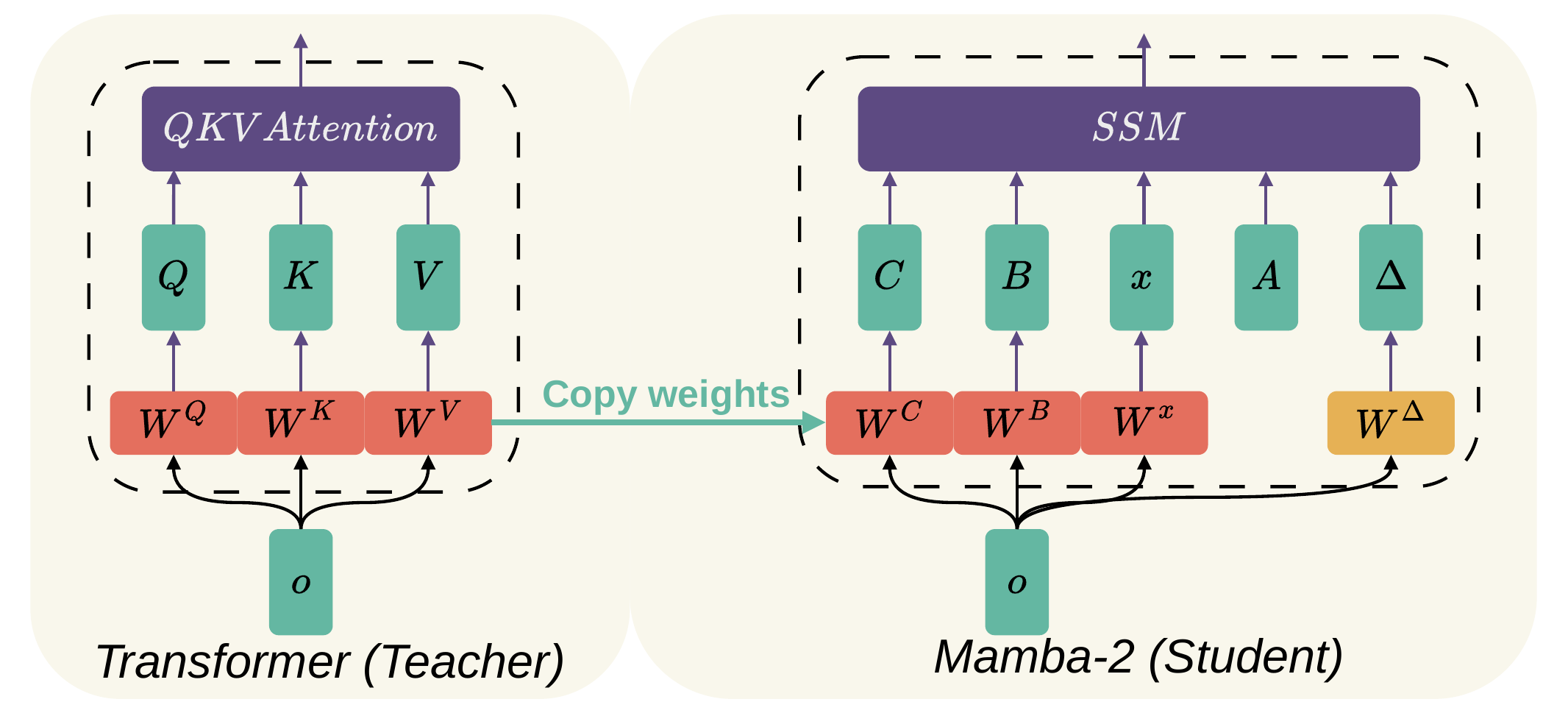}
    \caption{
    Following the alignment between Eq.~\ref{eq:ssm_discrete} and Eq.~\ref{eq:closed_form}, projection weights for $C$, $B$, and $x$ in Mamba are initialized with Transformer's $Q$, $K$, and $V$ projection weights, respectively.\looseness=-1
    }
    \label{fig:weight_init}
    \vspace{-4ex}
\end{figure}

\begin{figure*}[t]
    \centering
    \begin{minipage}[t]{0.36\textwidth}
        \centering
        \includegraphics[width=\linewidth]{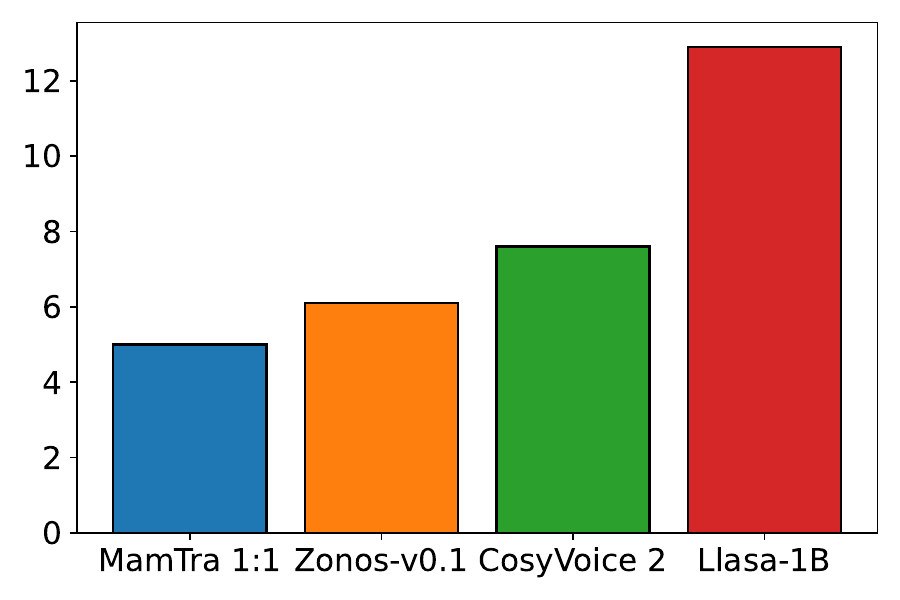}
        \caption{Memory usage (GB) on Seed-TTS-eval (using NVIDIA A6000). \algo~1:1 reduces VRAM by 34\% and 17\% compared to CosyVoice~2 and Zonos-v0.1, respectively.}
        \label{fig:real_memory_usage}
    \end{minipage}
    \hspace{0.02\textwidth} 
    \begin{minipage}[t]{0.59\textwidth}
        \centering
        \includegraphics[width=\linewidth]{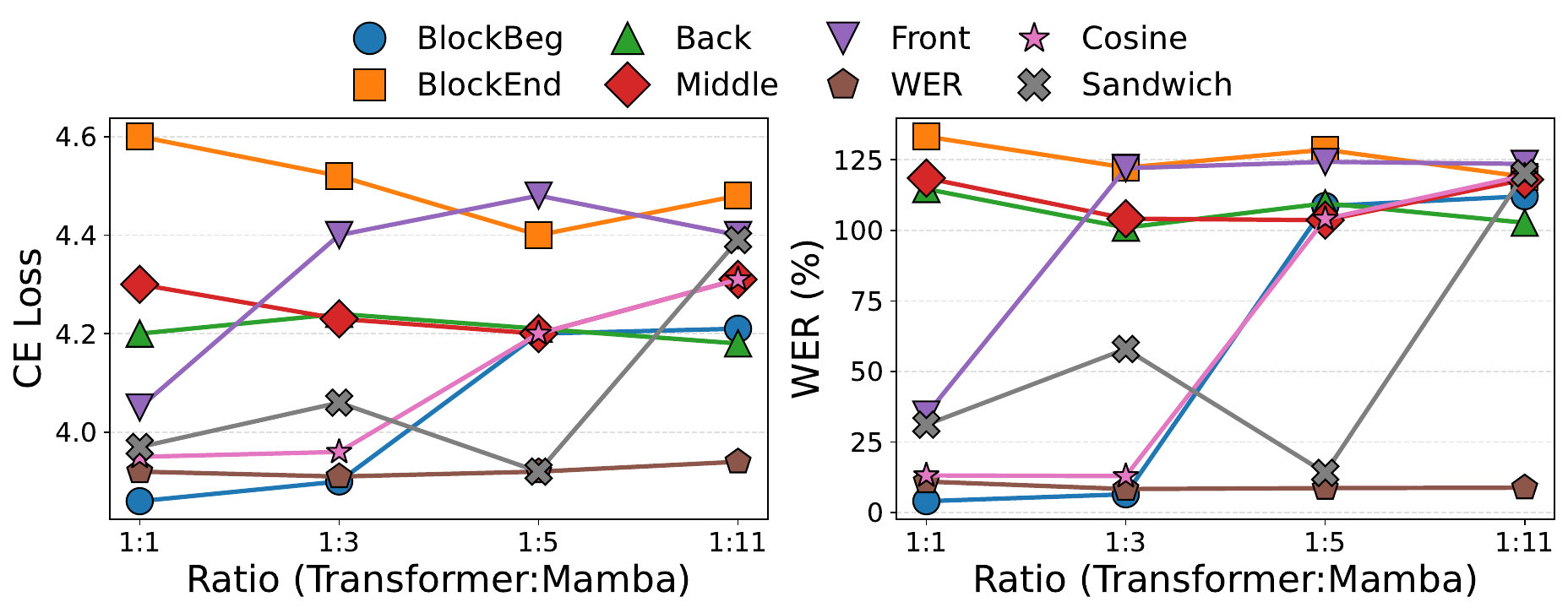}
        \caption{Cross-entropy (CE) loss and Word Error Rate (WER) on Seed-TTS-eval for hybrid Mamba-Transformer variants after 15 training epochs on LibriTTS (0.5k h). CE loss exhibits a consistent correlation with WER across different ratios and placement strategies.}
        \label{fig:ce_wer}
    \end{minipage}
    \vspace{-4ex}
\end{figure*}

\begin{figure}[t]
    \centering
    \includegraphics[width=\linewidth]{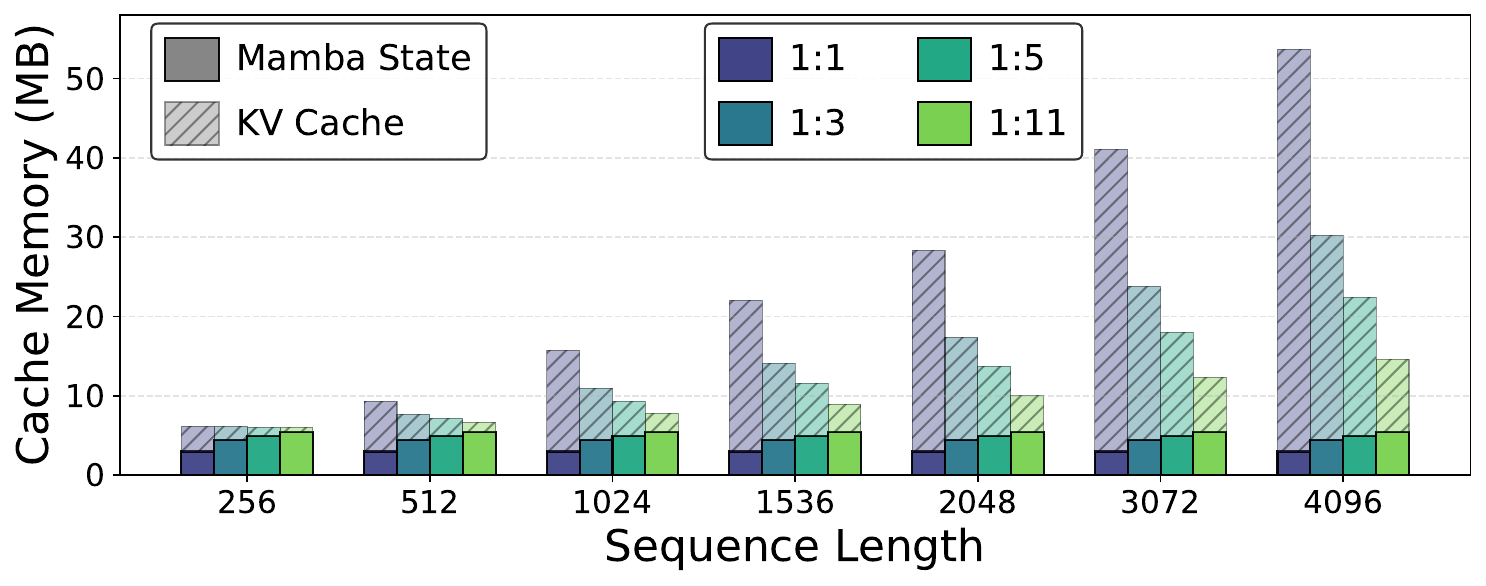}
    \caption{Cache size growth in the hybrid model, where the sequence-length dependency scales with the number of remaining Transformer layers.}
    \label{fig:cache_memory}
    \vspace{-4ex}
\end{figure}

\subsection{Linearization and Structural Mapping}
The transition from self-attention to the recurrent form can be interpreted as a linearization~\cite{Katharopoulos2020transformers}. 
To bridge these architectures at timestep $t$, we apply a causal mask $m_{s,t} = 1$ for $s \le t$ ($0$ otherwise). Removing the softmax allows using associativity to decouple the query from historical context:\looseness=-1
\begin{equation}
    \mathbf{y}_t = \frac{\mathbf{Q}_t^\top}{\sqrt{d_k}} \left( \sum_{s \le t} m_{s,t} \mathbf{K}_s \mathbf{V}_s \right).
    \label{eq:linear_attn}
\end{equation}
This formulation reveals an underlying recurrence where the accumulated product $\mathbf{h}_t = \sum_{s \le t} m_{s,t} \mathbf{K}_s \mathbf{V}_s$ serves as the latent state. Utilizing the property $m_{s,t} = m_{t-1,t} \cdot m_{s,t-1}$ for causal sequences (i.e., enabling $\mathbf{h}_t$ to be updated recursively from $\mathbf{h}_{t-1}$), the update rule is expressed in a linear RNN form:\looseness=-1
\begin{equation}
    \mathbf{h}_t = m_{t-1,t} \mathbf{h}_{t-1} + \mathbf{K}_t \mathbf{V}_t, \quad \mathbf{y}_t = \frac{1}{\sqrt{d_k}} \mathbf{Q}_t^\top \mathbf{h}_t.
    \label{eq:closed_form}
\end{equation}
Comparing Eq.~\ref{eq:ssm_discrete} and Eq.~\ref{eq:closed_form} reveals a direct structural alignment. Specifically, the key-value product ($\mathbf{K}_t \mathbf{V}_t$) corresponds to the SSM's input update ($\mathbf{\overline{B}}_t \mathbf{x}_t$), with the key ($\mathbf{K}_t$) mapping to the input-dependent parameter $\mathbf{B}_t$ and the value ($\mathbf{V}_t$) to the projected input $\mathbf{x}_t$. The query ($\mathbf{Q}_t$) aligns perfectly with the output projection ($\mathbf{C}_t$). Leveraging this equivalence, the pretrained Transformer's query, key, and value projection weights are directly transferred to initialize the corresponding Mamba layers for projections of $\mathbf{C}_t$, $\mathbf{B}_t$, and $\mathbf{x}_t$ (Figure~\ref{fig:weight_init}).
\looseness=-1

\subsection{Multi-Level Distillation}
While the closed-form weight transfer provides a strong structural initialization, relying solely on these weights is insufficient to preserve the teacher's performance. Specifically, removing the softmax nonlinearity fundamentally alters the attention dynamics, thereby reducing the model's expressive capacity~\cite{wang2024mambainllama}. To bridge this performance gap, MamTra employs a multi-level distillation strategy designed to transfer both the representational structure and the generation behavior of the teacher Transformer into the hybrid student model. The overall training objective is defined as:\looseness=-1
\begin{equation}
\mathcal{L} = \mathcal{L}_{\mathrm{CE}} + \mathcal{L}_{\mathrm{logits}} + \mathcal{L}_{\mathrm{emb}},
\end{equation}
where $\mathcal{L}_{\mathrm{CE}}$ provides ground-truth output supervision via cross-entropy. To recover the generation behavior lost during linearization, $\mathcal{L}_{\mathrm{logits}}$ minimizes the skew KL divergence~\cite{kodistillm} between the teacher's and student's logits. Finally, $\mathcal{L}_{\mathrm{emb}}$ enforces a mean-squared error constraint on the token embeddings per input sequence. Together, these objectives align the student's representations with the teacher's joint semantic-acoustic space, ensuring stable adaptation to the Mamba blocks despite the architectural substitution.\looseness=-1

\section{Experimental Setup}

\textbf{Datasets.}\,\,\,\,We utilize the LibriTTS dataset~\cite{zen2019libritts} for model training, and the Seed-TTS-eval \texttt{test-en} benchmark~\cite{anastassiou2024seed} alongside the LibriTTS \texttt{test-clean} split for evaluation. While Seed-TTS-eval \texttt{test-en} spans a balanced target text length ranging from 3 to 27 words, we additionally construct a LibriTTS-based test set encompassing a significantly wider length range (1 to 62 words). This evaluates the robustness of \algo~under length-stressed conditions.\looseness=-1

\vspace{2pt}
\noindent \textbf{Baselines.}\,\,\,\,We compare our approach against three strong baselines: (1) \textbf{CosyVoice~2}~\cite{du2024cosyvoice2}, built on the Qwen2.5 architecture~\cite{hui2024qwen2}; (2) \textbf{Llasa-1B}~\cite{ye2025llasa}, utilizing Llama-3.2 \cite{dubey2024llama}; and (3) \textbf{Zonos-v0.1}~\cite{zyphra2025zonos}, which represents the first hybrid Transformer-SSM model for TTS, although its precise architectural details remain largely unreported.\looseness=-1

\vspace{2pt}
\noindent \textbf{Training Details.}\,\,\,\,We conduct the performance recovery and fine-tuning of \algo~based on the CosyVoice 2 backbone \cite{du2024cosyvoice2}. All models are trained on NVIDIA A6000 GPUs using the Adam optimizer with a learning rate of $1e^{-5}$ and a dynamic batch size of 40,000 tokens.
While most experimental configurations are trained for 15 epochs to analyze architectural trends, models for the final rigorous comparisons (Table~\ref{tab:tts-comparison}) are trained for 50 epochs to ensure full convergence.\looseness=-1

\vspace{2pt}
\noindent \textbf{Evaluation Metrics.}\,\,\,\,We comprehensively assess the models based on computational efficiency, alongside the intelligibility and perceptual quality. Computational efficiency is quantified using TFLOPs, GPU memory usage, and cache size. For intelligibility, we report the Word Error Rate (WER). Perceptual quality is assessed via Speaker Similarity (SSIM), UTMOS, and Naturalness Mean Opinion Score (NMOS). All objective metrics are computed utilizing the \texttt{versa} toolkit~\cite{shi2025versa}. For the subjective NMOS evaluation, 50 utterances are randomly sampled per model and rated by 15 independent listeners.\looseness=-1

\begin{table}[t]
    \centering
    \addtolength{\tabcolsep}{3pt}
    \caption{Comparison of sequence-scaling computational and memory complexity across architectures. Here, $L$ denotes sequence length, $d$ the hidden dimension, $d_{kv}$ the key-value dimension in GQA~\cite{ainslie2023gqa} , and $N$ the state size of Mamba.}
    \label{tab:complexity_comp}
    \resizebox{\columnwidth}{!}{
    \begin{tabular}{l|c|cc}
    \toprule
    \multirow{2.5}{*}{\textbf{Model}} 
    & \textbf{Compute} 
    & \multicolumn{2}{c}{\textbf{Memory}} \\
    \cmidrule(lr){2-2} \cmidrule(lr){3-4}
    & {\textbf{FLOPs}} 
    & {\textbf{Training}} 
    & {\textbf{Inference}} \\
    \midrule
    {MHA} 
    & $\mathcal{O}(L^2 d + L d^2)$ 
    & $\mathcal{O}(L^2 + L d)$ 
    & $\mathcal{O}(L d )$ \\
    {GQA} 
    & $\mathcal{O}(L^2 d + L d^2)$ 
    & $\mathcal{O}(L^2 + L d)$ 
    & $\mathcal{O}(L d_{kv})$ \\
    {Mamba} 
    & $\mathcal{O}(L d^2)$ 
    & $\mathcal{O}(L d)$ 
    & $\mathcal{O}(d N)$ \\
    \bottomrule
    \end{tabular}
    }
    \vspace{-3ex}
\end{table}

\begin{table*}[t]
\centering
\addtolength{\tabcolsep}{-3pt}
\caption{Comparison of TTS Models Across Quality and Efficiency Metrics in {\textit{Seed-TTS-eval \texttt{test-en}}} and {\textit{LibriTTS \texttt{test-clean}}}. We compute the FLOPs needed for the LLM backbone to infer one token with a context length of 2048 ($1$ TFLOP $=$ $1e12$ FLOPs). All default cache enabled. NMOS is computed with confidence interval of $95\%$.}
\label{tab:tts-comparison}
\resizebox{\textwidth}{!}{
\begin{tabular}{l|c|c|c|c|c|c|c|c|c|c|c|c|c|c|c}
\toprule

\multirow{2.5}{*}{\textbf{Model}} &
\multirow{2.5}{*}{\textbf{Hybrid}} &
\multirow{2.5}{*}{\textbf{Source}} &
\multirow{2.5}{*}{\textbf{Params}} &
\multirow{2.5}{*}{\textbf{Layers}} &
\multirow{2.5}{*}{\textbf{Data\,($k$h)}} &
\multirow{2.5}{*}{\textbf{Ratio}} &
\multirow{2.5}{*}{\textbf{Strategy}} &
\multirow{2.5}{*}{\textbf{TFLOPs}\,$\downarrow$} &
\multirow{2.5}{*}{\textbf{NMOS}\,$\uparrow$} &
\multicolumn{3}{c|}{\textbf{\textit{Seed-TTS-eval \texttt{test-en}}}} &
\multicolumn{3}{c}{\textbf{\textit{LibriTTS \texttt{test-clean}}}} \\
\cmidrule(lr){11-13} \cmidrule(lr){14-16}
& & & & & & & & & &
\textbf{UTMOS\,$\uparrow$} & \textbf{SSIM\,$\uparrow$} & \textbf{WER\,$\downarrow$} &
\textbf{UTMOS\,$\uparrow$} & \textbf{SSIM\,$\uparrow$} & \textbf{WER\,$\downarrow$} \\
\midrule

Groundtruth & -- & -- & -- & -- & -- & -- & -- &
-- & 
$3.93\pm0.15$ & 
$3.52$ & 
$1.00$ & 
$1.47$ & 
$4.14$ & 
$1.00$ & 
$1.85$ \\ 
\midrule

Llasa-1B & $\times$ & Public & $1.3$B & $16$ & $250$ & $1{:}0$ & -- & 
$4.53$ & 
$3.64 \pm0.19$ & 
$4.13$ & 
$0.46$ & 
$3.54$ & 
$4.30$ & 
$0.40$ & 
$3.07$ \\ 

CosyVoice 2 & $\times$ & Public & $0.5$B & $24$ & $ 170$ & $1{:}0$ & -- & 
$1.78$ & 
$3.68\pm0.16$ & 
$\smash{{4.15}}$ & 
$\smash{{0.66}}$ & 
$\smash{{2.03}}$ & 
$\smash{{4.35}}$ & 
$\smash{{0.74}}$ & 
$\smash{{2.04}}$ \\ 
\midrule

Zonos-v0.1 & $\checkmark$ & Public & $1.6$B & $46$ &  $ 200$ & $1{:}3$ & BlockBeg & 
$7.32$ & 
$3.18\pm0.18$ & 
$3.63$ & 
$\smash{{0.67}}$ & 
$3.42$ & 
$3.96$ & 
$0.72$ & 
$3.37$ \\ 

\midrule
\rowcolor{gray!20} \cellcolor{white}& $\checkmark$ & Finetuned & $0.5$B &  $24$ & $ 0.5$ & $1{:}1$ & BlockBeg & 
$1.64$ & 
$3.66\pm0.16$ & 
${4.16}$ & 
${0.72}$ & 
$\smash{{2.28}}$ & 
$4.35$ & 
$\smash{{0.75}}$ & 
$\smash{{2.26}}$ \\ 
\cellcolor{white} & $\checkmark$ & Finetuned & $0.5$B  & $24$ & $ 0.5$ & $1{:}3$ & BlockBeg & 
$1.57$ & 
$3.65\pm0.16$ & 
$4.14$ & 
$0.72$ & 
$3.26$ & 
$4.35$ & 
$0.75$ & 
$2.99$ \\ 
\rowcolor{gray!20} \cellcolor{white} & $\checkmark$ & Finetuned & $0.5$B & $24$ & $ 0.5$ & $1{:}5$ & WER & 
$1.54$ & 
$3.65\pm0.16$ & 
$4.13$ & 
$0.72$ & 
$2.53$ & 
$4.36$ & 
$0.75$ & 
$2.28$ \\ 
\cellcolor{white} \multirow{-4}{*}{\textbf{\algo}} & $\checkmark$ & Finetuned & $0.5$B & $24$ & $ 0.5$ & $1{:}11$  & WER & 
$1.52$ & 
$3.27\pm0.18 $ & 
$4.15$ & 
$0.72$ & 
$3.99$ & 
$\smash{{4.37}}$ & 
$0.75$ & 
$4.08$ \\ 
\bottomrule
\end{tabular}}
\vspace{-3ex}
\end{table*}

\begin{table}[t]
\footnotesize
\caption{Impact of different loss components after 15 epochs.}
\label{tab:loss-ablation}
\resizebox{0.45\textwidth}{!}{
\begin{tabular}{l|c|c|c|c}
\toprule
\textbf{Components} & \textbf{WER} $\downarrow$ & \textbf{CER} $\downarrow$ & \textbf{SSIM} $\uparrow$ & \textbf{UTMOS} \\
\midrule
\algo~1:1 &    
$3.48$ &
$2.86$ &
$0.72$ &
$4.16$
\\
\midrule
\,\,$w/o$ $\gL_{CE}$     &    
$6.70$  &
$4.32$ &
$0.72$ &
$4.15$
\\
\,\,$w/o$ $\gL_{logits}$  &  
$6.13$  &   
$3.46$ &
$0.72$ &
$4.15$
\\
\,\,$w/o$ $\gL_{emb}$     &
$5.63$  &      
$3.26$ &
$0.72$ &
$4.15$
\\
\bottomrule
\end{tabular}}
\vspace{-4ex}
\end{table}

\section{Analysis}

\subsection{Efficiency and Complexity Analysis}

The structural mapping between Attention and Selective SSMs yields significant scaling differences. As shown in Table~\ref{tab:complexity_comp}, while GQA-based backbones like Llasa and CosyVoice 2 reduce cache footprint, they retain quadratic $O(n^{2}d)$ complexity. Conversely, Mamba~\cite{dao2024transformers} achieves linear $O(nd^{2})$ scaling. For \textbf{\algo}, this hybrid integration enables \textit{sub-quadratic computation} and \textit{sub-linear cache growth}, as complexity is dominated by the remaining Transformer layers. This is empirically confirmed in Figure~\ref{fig:cache_memory}, where Mamba state remains nearly constant while KV-cache growth is significantly attenuated relative to sequence length. In practice, \algo~reduces average inference-time memory by up to 34\% compared to the CosyVoice 2 baseline, as demonstrated in Figure~\ref{fig:real_memory_usage}.

\subsection{Efficiency-Quality Tradeoff}

As shown in Figure~\ref{fig:ce_wer}, the BlockBeg scheme consistently achieves lower CE and WER across ratios, while alternative placements exhibit higher variance and reduced robustness as the Mamba ratio increases. Notably, as the degree of layer removal becomes more aggressive, the advantage of WER-based layer importance becomes increasingly pronounced, consistent with prior observations in structured pruning~\cite{nguyen2025spade}. Based on these consistent findings under low-cost settings, we select the \textbf{BlockBeg} strategy as the replacement scheme for ratios of 1:1 and 1:3; \textbf{WER} strategy for ratios 1:5 and 1:11.

Table~\ref{tab:tts-comparison} illustrates \algo's quality--efficiency trade-off relative to Transformer-only and prior hybrid models. While the backbone network (\textbf{CosyVoice 2}) provides high quality, it incurs significant attention-based overhead. By replacing attention layers with Mamba blocks, \algo~reduces per-token FLOPs by up to $1.4\times10^{11}$ at a 2048 context length while maintaining comparable NMOS, UTMOS, and speaker similarity. Specifically, the \textbf{\algo~1:1} configuration preserves strong intelligibility, matching the teacher's perceptual metrics with only a modest $0.25\%$ absolute WER increase. Other variants, such as 1:3 and 1:5, offer further computational gains with manageable increases in word error rate. However, the most aggressive 1:11 ratio represents the upper bound of this trade-off, where intelligibility begins to degrade below the Zonos-v0.1 baseline. Furthermore, \algo~achieves up to a 34\% reduction in average inference memory, consistent with the sub-linear cache scaling in Figure~\ref{fig:cache_memory}. These results demonstrate that \algo~effectively balances naturalness with the efficiency required for memory-constrained, long-context TTS deployment.

\subsection{Sensitivity Under Length-Stress Conditions}
In the length-stress condition (1–62 words), the \algo~1:5 (WER) configuration demonstrates a remarkably stable efficiency-quality trade-off. Despite the aggressive substitution of attention layers for Mamba blocks, the importance-based WER selection strategy keeps the WER exceptionally low at 2.28\%. Notably, this configuration exhibits superior robustness compared to the more conservative 1:3 (BlockBeg) variant (WER $= 2.99\%$) and achieves parity with the 1:1 (BlockBeg) baseline while maintaining high UTMOS of $4.36$ and SSIM of $0.75$, consistently surpassing Zonos.
This suggests that under length-stressed conditions, better layer placement is more effective than simply retaining more Transformer blocks.\looseness=-1

\begin{figure}[t]
    \centering
    \includegraphics[width=0.95\linewidth]{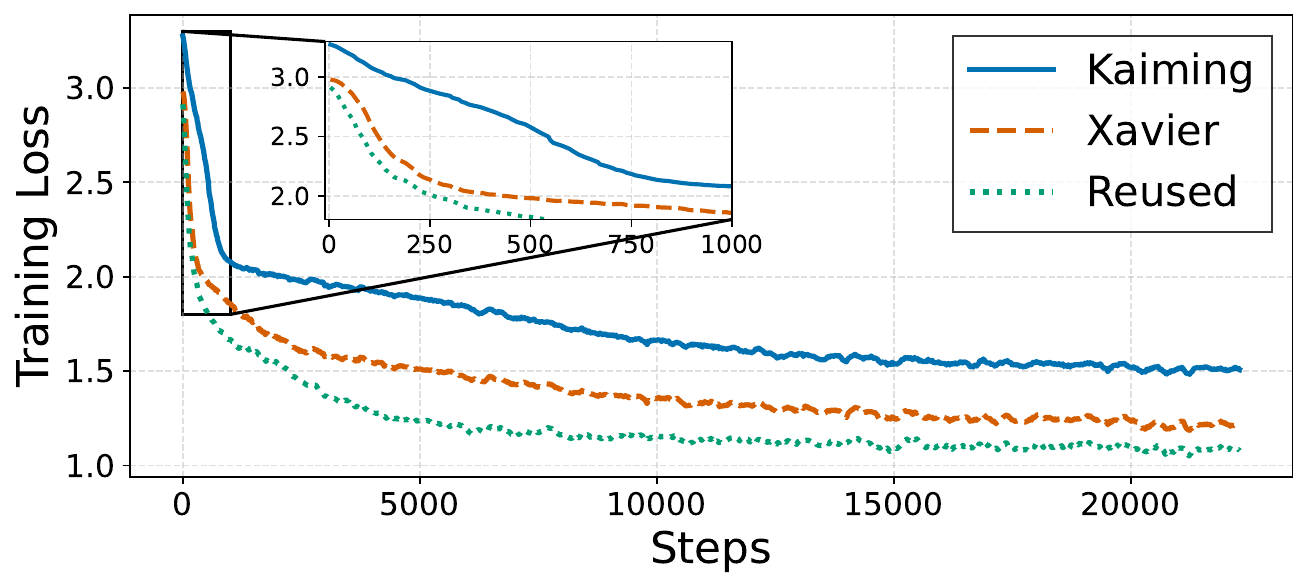}
    \caption{Convergence speed after 15 epochs.}
    \label{fig:weight_init_comparision}
    \vspace{-4ex}
\end{figure}

\subsection{Training Dynamics and Loss Contributions}
Ablation studies summarized in Table~\ref{tab:loss-ablation} confirm that $\mathcal{L}_{\mathrm{CE}}$, $\mathcal{L}_{\mathrm{logits}}$, and $\mathcal{L}_{\mathrm{emb}}$ are essential for effective knowledge transfer, primarily impacting linguistic accuracy rather than perceptual quality. Removing ground-truth supervision ($\mathcal{L}_{\mathrm{CE}}$) results in the most severe degradation in intelligibility, while the logit distillation loss ($\mathcal{L}_{\mathrm{logits}}$) using Skew KL-Divergence is critical for recovering the teacher's nuanced generation behavior.

Complementary to these objectives, the initialization strategy significantly impacts training efficiency. As illustrated in Figure~\ref{fig:weight_init_comparision}, reusing pretrained Transformer weights for both attention and Mamba blocks facilitates much faster convergence and a lower final training loss compared to standard Xavier or Kaiming initialization. This ``initialize-and-train" approach allows the hybrid model to maintain a clear performance gap from the outset, enabling successful performance recovery using less than 2\% of the original teacher's training data.

\section{Conclusion}
We proposed \algo, a hybrid Mamba-Transformer framework designed for efficient and high-quality speech synthesis.
Through rigorous analysis, we optimized the integration and introduced custom initialization with multi-level distillation. Our experimental results demonstrate that \algo~achieves up to a 34\% reduction in inference memory and saves $1.4e^{11}$ FLOPs per token without compromising generation quality. These findings suggest that \algo~serves as a practical and scalable solution for memory-efficient TTS systems. We hope that our comprehensive analysis provides a solid foundation for future research in hybrid modeling for speech tasks.\looseness=-1

\clearpage
\newpage
\section{Acknowledgement}
This work was supported by the IITP-ITRC grant funded by the Korea government (Ministry of Science and ICT, IITP-2026-RS-2023-00259991).

\section{Generative AI Use Disclosure}
During the preparation of this manuscript, the authors used Google's Gemini as a writing assistant for editing, language polishing, and structural refinement. This tool was used solely to improve the readability and overall clarity of the prose. For implementation-level assistance, the authors utilized xAI's Grok for code refinement and debugging, and Anthropic's Claude for code cleaning ahead of public release and for project page creation. These tools were not involved in the design of the proposed methods, experimental protocols, or scientific interpretations. All content, code, and conclusions were reviewed, revised, and approved by the authors, who take full responsibility for the integrity of the manuscript and its submission.
\bibliographystyle{IEEEtran}
\bibliography{shortstrings,mybib}

\end{document}